\documentclass{moriond}

\def\Journal#1#2#3#4{{#1} {\bf #2}, #3 (#4)}

\def\NPB{{\em Nucl. Phys.} B}

\def\PRD{{\em Phys. Rev.} D}

\def\be{\begin{equation}}
\def\ee{\end{equation}}
\def\bea{\begin{eqnarray}}
\def\eea{\end{eqnarray}}

\newcommand{\Photo}{}

\begin{document}
\vspace*{4cm}
\title{Lepton flavour universality tests in $b \to c l \nu$ decays at the LHCb experiment}
\author{Abhijit Mathad, On behalf of the LHCb collaboration}
\address{Physik-Institute, University of Zurich \\Winterthurerstrasse, 8057 Zürich, Switzerland}
\maketitle\abstracts{Semileptonic charged current $b \to c l^- \overline{\nu}_l$ decays offer a compelling avenue to probe the limits of the Standard Model and investigate lepton flavor universality (LFU). While the SM maintains flavor universality, new physics models can introduce interactions that violate this universality, particularly for decays involving third-generation quarks and leptons. An enticing observable to test LFU is the ratio of branching fractions of $b \to c \tau^- \overline{\nu}_\tau$ and $b \to c \mu^- \overline{\nu}_\mu$ decays. Recently, the LHCb experiment measured this ratio using $\overline{B} \to D^{(*)} l^- \overline{\nu}_l$ decays. We discuss these results in this proceedings of the 57th Rencontres de Moriond on Electroweak Interactions and Unified Theories.}
The semileptonic charged current $b \to c l^- \overline{\nu}_l$
decays are an excellent avenue for conducting rigorous tests of the Standard Model (SM). Previous studies have utilised these decays to explore various phenomena, such as testing the unitarity of the Cabibbo-Kobayashi-Maskawa (CKM) matrix, examining the production properties of $b$-hadrons and $c$-hadrons, investigating neutral $b$-meson mixing properties, and analysing decay properties like branching fractions and form factors. Additionally, these decays have been used to investigate lepton flavor universality, which is the central focus of this proceedings that follows the talk given at the 57th Rencontres de Moriond on Electroweak Interactions and Unified Theories.

In the Standard Model (SM), the couplings of the electroweak force to leptons are flavor universal, with the exception of the Yukawa couplings of the Higgs boson. However, new physics models can introduce interactions that violate lepton flavor universality, leading to an enhancement of the decay rate of one lepton flavor over another. These models predict that the decay rates related to the third generation of leptons will be particularly affected~\cite{np1,np2}.

One observable that is sensitive to such new physics effects is the lepton flavor universality (LFU) ratio, 
which is defined as the ratio of the branching fractions of $b \to c \tau^- \overline{\nu}_\tau$
and $b \to c \mu^- \overline{\nu}_\mu$ and is given by:
\begin{equation}
R(X_c) = \frac{\mathcal{BF}(X_b \to X_c \tau^- \overline{\nu}_\tau)}{\mathcal{BF}(X_b \to X_c \mu^- \overline{\nu}_\mu)},
\end{equation}
where $X_b$ and $X_c$ represent $b$ and $c$ mesons, respectively. 

The LFU ratio offers several advantages as an observable, primarily due to its theoretical cleanliness. In particular, the uncertainties associated with form factors and CKM matrix elements largely cancel out in this ratio. Additionally, the common systematic uncertainties related to detection efficiencies and yield estimation from data also cancel out. The LHCb experiment is well-suited to measuring this observable with high precision, owing to the large $b$-hadron production cross-section and sizeable branching 
fractions of $b \to c l^- \overline{\nu}_l$
decays, which are typically of the order of $10^{-2}$.
However, measuring the LFU ratio also presents several challenges. The presence of multiple missing neutrinos in the final state affects the resolution of the kinematic variables used to distinguish between signal and background events. Moreover, there is a significant contamination from partially reconstructed background events that must be modelled accurately. Furthermore, large simulation samples are required to model both the signal and background events.

Previously, both the LHCb and $B$ factories have measured the LFU ratio in 
$\overline{B} \to D^{(*)} l^- \overline{\nu}_l$
decays and found a deviation from the SM predictions at the level of $3.3\sigma$~\cite{hflav} (before 2023). 
Additionally, the LHCb experiment has measured the LFU ratio in 
$\overline{B} \to J/\psi l^- \overline{\nu}_l$
decays~\cite{rjpsi}, denoted as $R(J/\psi)$, and in 
$\Lambda_b \to \Lambda_c^+  l^- \overline{\nu}_l$
decays~\cite{rlc}, denoted as $R(\Lambda_c)$. 

Recently, the LHCb experiment conducted two new measurements of the LFU ratio in 
$\overline{B} \to D^{(*)} l^- \overline{\nu}_l$ decays.
The first measurement is a combined measurement of $R(D)$ and $R(D^)$, and is discussed in detail in Section~\ref{sec:muonic}. The second measurement is the measurement of $R(D^*)$ using hadronic $\tau$ decays, and is discussed in detail in Section~\ref{sec:hadronic}.

\section{Combined measurement of $R(D)$ and $R(D^*)$ with muonic $\tau$ decays}
\label{sec:muonic}

The LHCb experiment has performed a combined measurement of $R(D)$ and $R(D^*)$~\cite{newrdst}, 
utilising the large branching fraction of approximately 17.4\% of muonic 
$\tau^- \to \mu^- \overline{\nu}_{\mu} \nu_{\tau}$ decays in the final state. The analysis uses the Run 1 data sample corresponding to an integrated luminosity of $3 \mathrm{fb^{-1}}$, with proton-proton collisions at a center-of-mass energy of $7$ and $8$ TeV. This measurement updates a previous analysis~\cite{oldrdst} that used the same data sample to measure $R(D^*)$ alone. The current analysis has several improvements, including an improved modelling of signal and background shapes used in the measurement of the yields, and a reduction in the misidentification of hadrons as muons.

\begin{figure}[!htp]
 \centering
 \includegraphics[width=1.0\textwidth]{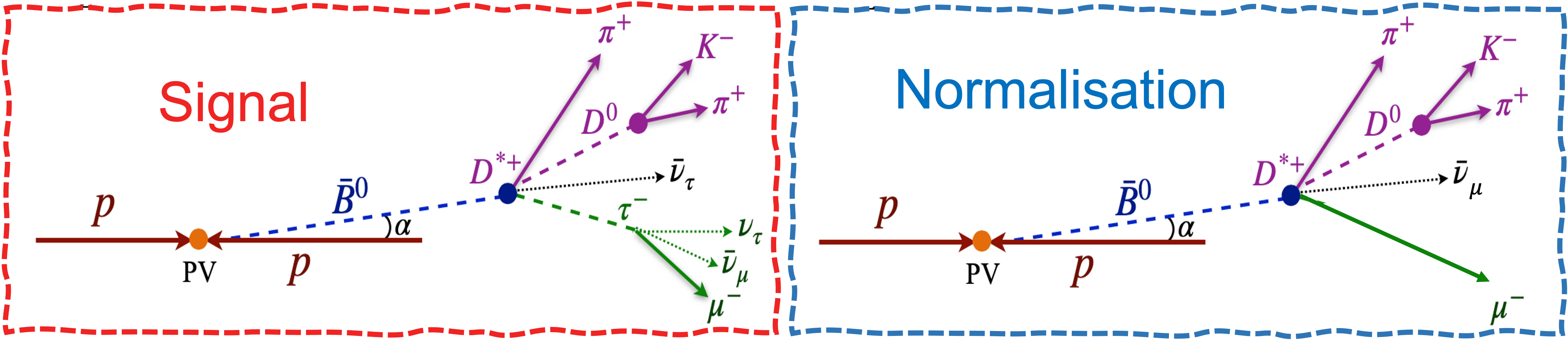}
\caption[]{
 Topology of one of the signal 
 $\overline{B}^0 \to D^{*+} \tau^- \overline{\nu}_{\tau}$  (left)
 and normalisation
 $\overline{B}^0 \to D^{*+} \mu^- \overline{\nu}_{\mu}$ (right) channels.
 }
 \label{fig:signal_topology}
\end{figure}

In this study, three signal channels are investigated:
$\overline{B}^0 \to D^{+} \tau^- \overline{\nu}_{\tau}$,
$B^- \to D^{*0} \tau^- \overline{\nu}_{\tau}$, and
$B^- \to D^0 \tau^- \overline{\nu}_{\tau}$, with the topology of one of the signal channels
shown in Fig.~\ref{fig:signal_topology}.
To determine the yields of the three signal channels,
two disjoint samples are used.
One sample corresponds to $D^{*+} \mu^-$, which includes the contribution from the signal $\overline{B}^0 \to D^{+} \tau^- \overline{\nu}_{\tau}$ decays.
The other sample corresponds to a large sample of $D^{0} \mu^-$, where the $D^{*+}$ contribution is vetoed,
and includes the contributions from all three signal channels.

The $\overline{B} \to D^{(*)} \mu^- \overline{\nu}_{\mu}$ channel serves as the normalisation for measuring $R(D^{(*)})$ due to its similar decay topology as the signal channel (see Fig.~\ref{fig:signal_topology}), 
allowing for cancellation of many common systematic uncertainties related to the detection efficiencies.
Since the normalisation channel is 20 times larger than the signal channels, precise resolution of kinematic observables is necessary to differentiate between them.

The missing neutrinos in the final state of the signal channels, poses a challenge
in the accurate reconstruction of the $B$ momentum. 
The transverse component of the $B$ momentum is obtained using the 
momentum of the $D^{(*)}\mu$ visible system and the direction of the flight distance vector.
The longitudinal component of the $B$ momentum is then determined using boost-approximation~\cite{nureco}. 
This allows to reconstruct the $B$ momentum with a resolution of about 20\%. 
Three kinematic observables are used as the fit variables to distinguish between the signal and normalisation channels.
These are the $q^2$ defined as the invariant mass squared of the di-lepton system, 
the missing mass squared $m_{miss}^2$  and $E_l$ defined as the energy of the lepton in the $B$ rest frame.

The analysis considers various background sources. The ``feed-down" and ``double-charm" backgrounds arise from $B$ hadron decays to excited charmed mesons (i.e. $\overline{B} \to D^{**} (\to D^{(*)} X) l^- \overline{\nu}_l$)) 
and two charmed mesons (i.e. $\overline{B} \to D^{(*)} D^{(*)} (\to l X) X$), respectively. These are reduced by an MVA-based isolation requirement, and any residual contribution is modelled using simulation. 
The ``misidentified" background arises from $B$ decays to a charmed meson and a hadron (i.e. $\overline{B} \to D^{(*)} h X$) where the hadron is misidentified as a muon, which is reduced by a muon identification requirement and modelled using data. 
The final background type is the "combinatorial" background originating from random track and muon combinations used to build fake $B$ and $D^{(*)}$ candidates. This is reduced by applying a requirement on the $B$ vertex fit quality and is modelled using same-sign data samples (i.e. $D^{(*+)} \mu^+$ and $D^0 \pi^- \mu^-$).

\begin{figure}[!htp]
 \centering
 \includegraphics[width=0.7\textwidth]{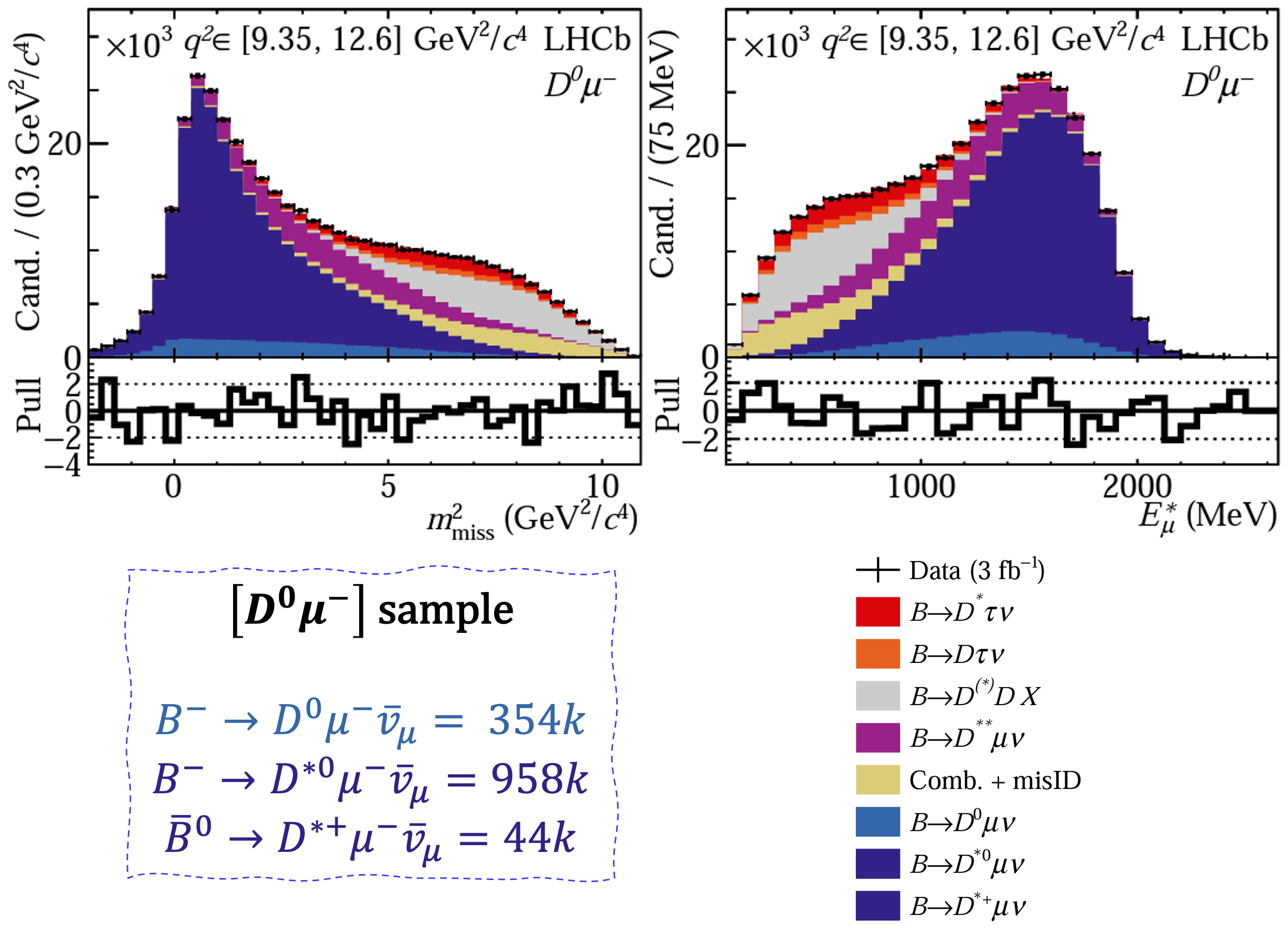}
\caption[]{
 Fit to $D^0\mu^-$ data samples in the highest $q^2$ bin. The yields of the involved semimuonic $B$ decays are also shown. 
 The fit to the $D^{*+}\mu^-$ data samples is shown in the Ref~\cite{newrdst}.
 }
 \label{fig:signal_norm_fit}
\end{figure}

To extract signal and normalisation yields, a binned maximum likelihood fit is performed to three kinematic observables: $q^2$, $m_{miss}^2$, and $E_l$. Eight data samples are fitted simultaneously, with each $D^{*+}\mu^-$ and $D^0\mu^-$ sample consisting of a fit to the signal region and three control regions. 
In the fit the signal shape for $B \to D^{(*)} l^- \overline{\nu}_l$ decays is modelled using 
Boyd, Grinstein, and Lebed (BGL) parametrization~\cite{bgl},
whereas the $B \to D^0 l^- \overline{\nu}_l$ decays are modelled using Bourrely, Caprini, and Lellouch (BCL) parametrization~\cite{bcl}.
Form factor parameters are inferred from the fit, with only helicity suppressed terms being constrained from external inputs. 
The fit result to the $D^0\mu^-$ data samples in the highest $q^2$ bin is shown in Fig.~\ref{fig:signal_norm_fit}.

The three data control regions are defined for each $D^{*+}\mu^-$ and $D\mu^-$ sample by reversing the isolation requirement. One control region uses an extra charged pion track to constrain the form factor parameters related to light excited charm mesons. Another control region uses two oppositely charged pion tracks to constrain the feed-down shape and decay properties of heavy excited charm mesons. The third control region uses an extra charged kaon track to control the phase space modelling of the double-charm background.

The analysis yielded the following values for the $R(D^{(*)})$ observables:
\be 
 R(D^{*+}) = 0.281 \pm 0.018 (stat) \pm 0.024 (syst)
\ee
\be 
 R(D^0) = 0.441 \pm 0.060 (stat) \pm 0.066 (syst)
\ee
The first uncertainty is statistical and the second is systematic.
The dominant sources of systematic uncertainties are the limited size of the simulated samples used to model signal and background,
and the assumptions made in the modelling of the double charm background and form factor parameters related to the feed-down background.
Both of these uncertainties should reduce with more data. The correlation coefficient between $R(D^{*+})$ and $R(D^0)$ is found to be $\rho=-0.43$.

\section{Measurement of $R(D^{*+})$ using hadronic $\tau$ decays}
\label{sec:hadronic}

\begin{figure}[!htp]
  \centering
  \includegraphics[width=1.0\textwidth]{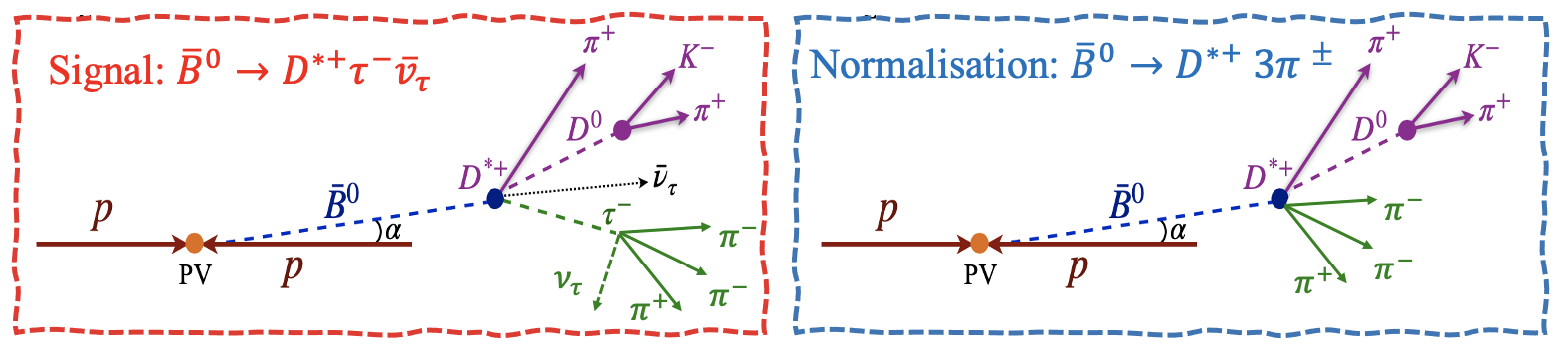}
\caption[]{
  The decay topology of the signal and normalisation channels.}
  \label{fig:hadronic_signal_norm}
\end{figure}

The LHCb collaboration has recently published a measurement of $R(D^{*+})$ using hadronic $\tau$ decays~\cite{newrdsthad}.
This analysis is based on a partial Run 2 data sample, corresponding to an integrated luminosity of 
$2\ \mathrm{fb^{-1}}$, and with proton-proton collisions at a center-of-mass energy of $13\ \mathrm{TeV}$. 
Compared to the Run 1 analysis~\cite{oldrdsthad}, we expect a two-fold increase in signal events, thanks to the higher center-of-mass energy and improved signal selection strategy. The analysis exploits the hadronic $\tau^- \to \pi^- \pi^+ \pi^- (\pi^0) \nu_{\tau}$ decay mode, which has a combined branching fraction of $13.5\%$. 
The measured quantity is the ratio of the branching fractions of $\overline{B}^0\to D^{*+} \tau^- \overline{\nu}_{\tau}$ and the normalisation channel 
$\overline{B}^0 \to D^{*+} \pi^- \pi^+ \pi^-$ (see Eq.~\ref{eq:rdst_had}), 
which has the same final state particles as the signal channel, leading to the cancellation of common systematic uncertainties,
\be 
  K(D^{*+}) = \frac{\mathcal{BF}(\overline{B}^0 \to D^{*+} \tau^- \overline{\nu}_{\tau})}{\mathcal{BF}(\overline{B}^0 \to D^{*+} \pi^- \pi^+ \pi^-)}.
\label{eq:rdst_had}
\ee
The decay topology of the signal and normalisation channels are shown in Fig.~\ref{fig:hadronic_signal_norm}. 
Using the external input of the ratio of the branching fractions of 
$\overline{B}^0 \to D^{*+} \pi^- \pi^+ \pi^-$ and 
$\overline{B}^0 \to D^{*+} \mu^- \overline{\nu}_{\mu}$~\cite{pdg}, we extract
\be
  R(D^{*}) = K(D^{*+}) \times \frac{\mathcal{BF}(\overline{B}^0 \to D^{*+} \pi^- \pi^+ \pi^-)}{\mathcal{BF}(\overline{B}^0 \to D^{*+} \mu^- \overline{\nu}_{\mu})}.
\ee

The analysis implemented a selection strategy to reduce background contributions, with the two most significant being ``prompt" ($B \to D^{*+} 3\pi X$)
and ``double-charm" ($B \to D^{+} D_s (\to 3 \pi X) X$) backgrounds.
The ``prompt" background was reduced by requiring the $\tau$ decay vertex downstream of the $B$ decay vertex, and a dedicated multivariate analysis using a boosted decision tree (BDT) classifier. 
The ``double-charm" background was reduced using 
an anti-$D_s$ BDT classifier trained on observables based on kinematics and resonant structures contributing to the $D_s \to 3\pi$ final state.
This anti-$D_s$ is also used an one of the fit variables 
to extract the signal yield.

After the full signal selection, the $B \to D^{+} D_s (\to 3 \pi) X$ channel form the largest background and hence
its accurate modelling is crucial for the analysis.
Firstly, the fractions of various resonant contributions to the $D_s \to 3\pi$ decay were obtained from a fit to the control region, 
which was defined by inverting the anti-$D_s$ BDT classifier.
The fit variables used were $m(\pi^+\pi^+)$, $min([m(\pi^+\pi^-)])$, $max[m(\pi^+\pi^-)]$ and $m(3\pi)$.
The results of this fit were used to correct the simulation sample used to model the $D_s \to 3\pi$ decay.
Secondly, to obtain the knowledge of various $B$ decay modes contributing to the $D_s$ production a fit to the $m(D^{+} 3\pi) - m(K\pi) - m(3\pi)$ distribution was performed using a data control region defined as $|m(D_s)| < 20\ \mathrm{MeV}$ and
removing the anti-$D_s$ BDT classifier requirement.
The result of this fit was a direct input to the signal fit.

\begin{figure}[!htp]
  \centering
  \includegraphics[width=0.8\textwidth]{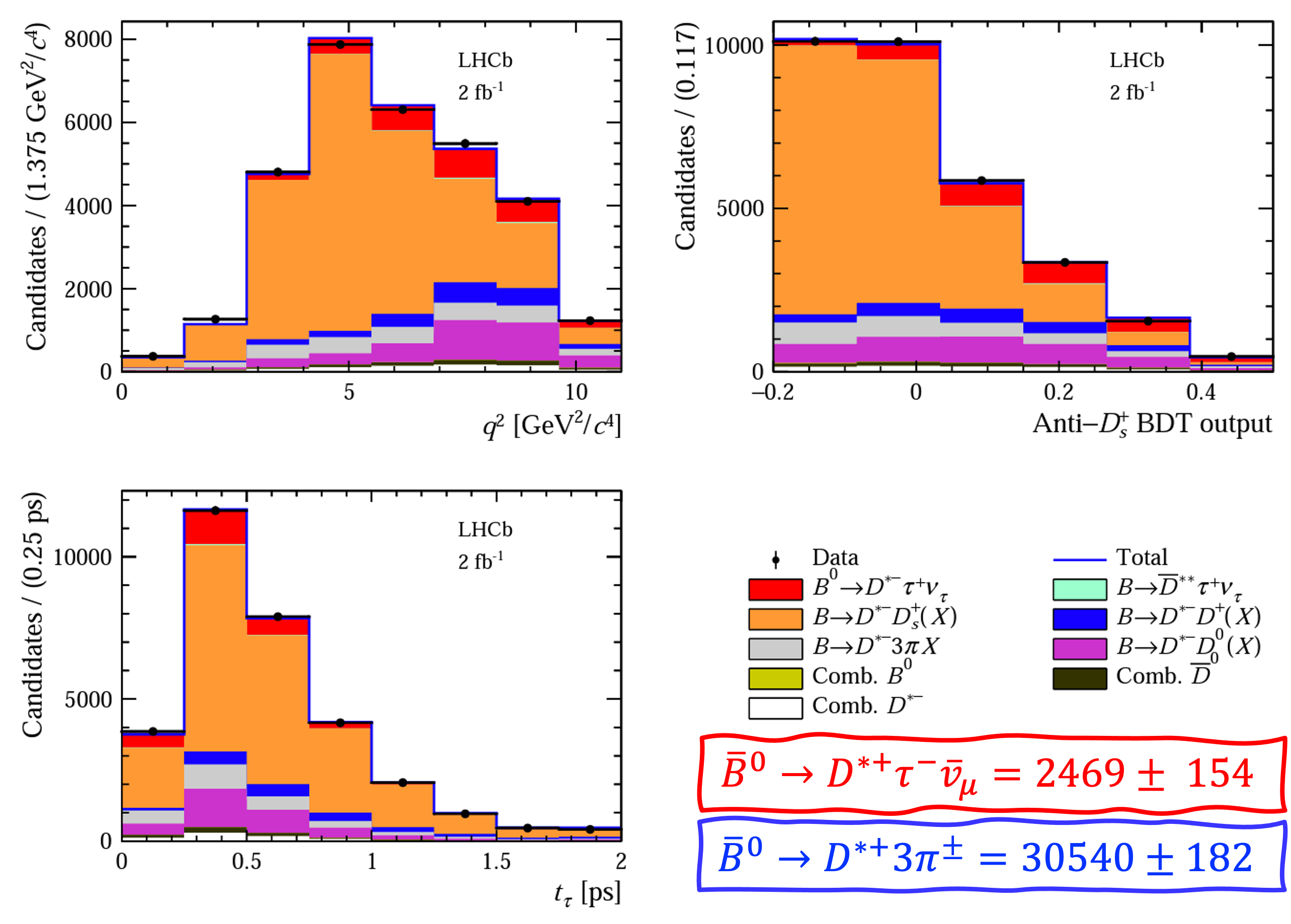}
\caption[]{
  Fit projections of the (top-left) $q^2$, (top-right) anti-$D_s$ BDT classifier output 
  and (bottom-left) $\tau$ lifetime.
  The signal and normalisation yields are also shown in the bottom-right plot.
  The fit projection of $m(D^{*+} \pi^- \pi^+ \pi^-)$ for the normalisation channel is also shown in Ref.~\cite{newrdsthad}.
  }
  \label{fig:hadronic_fit}
\end{figure}

To obtain the signal yield, a 3D maximum likelihood fit is conducted to the $q^2$, anti-$D_s$ BDT classifier output, and $\tau$ lifetime. The result of the fit is depicted in Fig.\ref{fig:hadronic_fit}, where the signal component is modelled using the Caprini, Lellouch and Neubert (CLN) parametrisation\cite{cln}. The normalisation yield is determined from a separate unbinned fit to the $m(D^{*+} \pi^- \pi^+ \pi^-)$ invariant mass. The signal and normalisation yields are also presented in Fig.~\ref{fig:hadronic_fit}.

The analysis results in the following measurement of the $K(D^*)$ ratio:
\be
K(D^*) = 1.70 \pm 0.10 (stat)^{+0.11}_{-0.10} (syst)\,
\ee
where the first uncertainty is statistical and the second is systematic.
The dominant sources of systematic uncertainty are the limited size of the simulation sample and
the limited knowledge of the background components' modelling.
Using external input of 
$\frac{\mathcal{BF}(\overline{B}^0 \to D^{*+} \pi^- \pi^+ \pi^-)}{\mathcal{BF}(\overline{B}^0 \to D^{*+} \mu^- \overline{\nu}_{\mu})}$, the $R(D^{*+})$ is measured to be:
\be
R(D^{*+}) = 0.247 \pm 0.015 (stat) \pm 0.015 (syst) \pm 0.012 (ext).
\ee
Here the first uncertainty is statistical, the second is systematic and the third is due to external inputs.
When combined with the Run 1 result, the $R(D^*)$ is measured to be:
\be 
R(D^{*+}) = 0.257 \pm 0.012 (stat) \pm 0.014 (syst) \pm 0.012 (ext).
\ee

\section{Summary and conclusions}

\begin{figure}[!htp]
  \centering
  \includegraphics[width=0.6\textwidth]{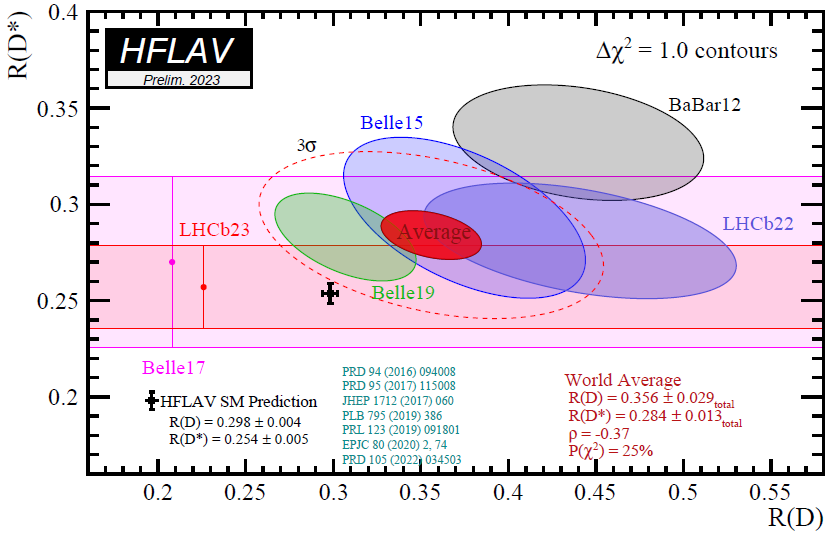}
\caption[]{
  The state-of-art related to $R(D)$ and $R(D^*)$ ratios, where the 
  world average shows a $3.2\sigma$ tension with the Standard Model prediction.
  }
  \label{fig:summary}
\end{figure}

The study of semileptonic charged current decays provides a powerful tool to investigate the universality of lepton flavor. 
In this Moriond proceedings, we have presented the two most recent measurements of the $R(D)$ and $R(D^)$ ratios~\cite{newrdst,newrdsthad}.
The results are in agreement with the previous measurements and the 
world average exhibits a $3.2\sigma$ tension~\cite{hflav} with the SM prediction (see Fig.~\ref{fig:summary}).
Further exploration of the LFU ratios in various $b$-hadron species will provide a deeper understanding of the potential new physics that may be responsible for the observed deviations. 
The upcoming experiments such as Belle II and LHCb Upgrade I will play a crucial role in this regard. 
Moreover, the ongoing theoretical developments in the SM predictions for these ratios will be important in interpreting the future experimental measurements. Thus, the continued efforts in both the theoretical and experimental fronts are necessary to uncover any hints of new physics beyond the SM in the lepton flavour universality sector.

\end{document}